\documentclass[10pt, conference, compsocconf]{IEEEtran}
\IEEEoverridecommandlockouts
\usepackage{graphicx}
\usepackage{multirow}
\usepackage{adjustbox}
\usepackage{amsmath}
\usepackage{cite}
\ifCLASSINFOpdf
\else
\fi
\begin{document}
\title{FM4SN: A Feature-Oriented Approach to Tenant-Driven Customization of Multi-Tenant Service Networks}

\author{\IEEEauthorblockN{Indika~Kumara,~Willem-Jan~van~den~Heuvel,~Damian~Tamburri}
\IEEEauthorblockA{Jheronimus Academy of Data Science\\Eindhoven University of Technology\\
's-Hertogenbosch, Netherlands\\
(i.p.k.weerasingha.dewage, wjheuvel, d.a.tamburri)@tue.nl}
\and
\IEEEauthorblockN{Jun~Han,~Alan~Colman}
\IEEEauthorblockA{School of Software and Electrical Engineering\\Swinburne University of Technology\\
Melbourne, Australia\\
(jhan, acolman)@swin.edu.au}
}


%Telephone: (800) 555--1212, Fax: (888) 555--1212}


\makeatletter
\def\ps@IEEEtitlepagestyle{
  \def\@oddfoot{\mycopyrightnotice}
  \def\@evenfoot{}
}
\def\mycopyrightnotice{
  {\footnotesize
  \begin{minipage}{\textwidth}
  \centering
\copyright2019 IEEE. Personal use of this material is permitted. Permission from IEEE must be obtained for all other uses, in any current or future media, including reprinting/republishing this material for advertising or promotional purposes, creating new collective works, for resale or redistribution to servers or lists, or reuse of any copyrighted component of this work in other works. DOI:10.1109/SCC.2019.00028
  \end{minipage}
  }
}

\maketitle
\IEEEpubidadjcol

\begin{abstract}
In a multi-tenant service network, multiple virtual service networks (VSNs), one for each tenant, coexist on the same service network. The tenants themselves need to be able to dynamically create and customize their own VSNs to support their initial and changing functional and performance requirements. These tasks are problematic for them due to: 1) platform-specific knowledge required, 2) the existence of a large number of customization options and their dependencies, and 3) the complexity in deriving the right subset of options. In this paper, we present an approach to enable and simplify the tenant-driven customization of multi-tenant service networks. We propose to use \textit{feature} as a high-level customization abstraction. A regulated \textit{collaboration} among a set of services in the service network realizes a feature. A software engineer can design a customization policy for a service network using the mappings between features and collaborations, and enact the policy with the controller of the service network. A tenant can then specify the requirements for its VSN as a set of functional and performance features. A customization request from a tenant triggers the customization policy of the service network, which (re)configures the corresponding VSN at runtime to realize the selected features. We show the feasibility of our approach with two case studies and a performance evaluation.
\end{abstract}


\IEEEpeerreviewmaketitle

\section{Introduction}
Business (IT) services are proxies for real world businesses of enterprises. Their capacities are usually constrained by the underlying business capabilities and product/service offerings rather just IT-infrastructure resources. They relate to each other and collaborate to realize business processes, forming a network of services connected according to their capabilities and interoperability \cite{R1, R2, R3}. The interactions between the services over the service network are regulated by the contractual relationships that exist between them. A service network can use the single instance multi-tenancy (SIMT) model \cite{R4} to share itself between different tenants to reduce operational cost of managing different service network variants for tenants, and improve utilization of the capacities of services \cite{R3}. The result is a \textit{multi-tenant service network}, where a set of virtual service networks (VSNs) simultaneously coexist on the same service network in a way analogous to virtual computer networks.

The provider of the composite application that realizes a multi-tenant service network can create and reconfigure VSNs for its tenants to support their functional and performance requirements. As the complexity of the application, the number of tenants, and the frequency of changes to their requirements increase, the cost of provisioning the application can also increase. To reduce this operational overhead, the provider can allow the tenants to create and reconfigure their own VSNs. However, these tasks are challenging for the tenants due to: 1) platform specific knowledge required, 2) large number of detailed options, and 3) difficulty of selecting consistent and mutually compatible options.

The customization support exists for classical composite applications \cite{R6, R7, R8} and multi-tenant applications \cite{R9, R10, R11, R12}. However, none considers the service network model. Service composition models such as process-centric models and component-based models fail to represent service networks naturally. Moreover, most works create and maintain physically separated application variants for tenants (the multi-instance multi-tenancy (MIMT) model). The customization of SIMT applications is limited functional requirements. In addition, these works do not sufficiently utilize the natural correspondence between domain requirement options and realization options to enable software engineers to design and execute customization decisions without undue complexity.

This paper presents \textit{FM4SN} (Feature Models for Service Networks), a tenant-oriented customization approach and framework for multi-tenant service networks. We base our work on the \textit{Software-Defined Service Networks (SDSN)} \cite{R3} approach that supports the design, enactment, and management of multi-tenant service networks. In SDSN, a service network (the application) is a runtime model (\textit{models@runtime}\cite{R5}) that acts as an overlay network over the services. With managed routing of message exchanges over the service network, VSNs can be dynamically formed. 

We use the concept of a \textit{feature} \cite{R13, R14} from software product lines to provide a suitable domain abstraction to model the commonality and variability in the customization requirements. A \textit{collaboration} among a subset of the services in the service network realizes a feature. A collaboration has a topology, and the message passing over this topology is regulated. The variations in services and their performance, topology, and regulations applied realize the performance variations of a functional feature. A feature-level change to a VSN alters the routing and regulation logic of the VSN at the granularity of a collaboration. To codify the customization decisions, we provide a rule-based policy language based on the mappings between features and collaborations. A software engineer can define the customization policy and the feature model for the service network, and deploy them in the manager of the service network. A tenant can make a customization request by selecting and un-selecting features and changing their properties. A customization request triggers the relevant policy rules, which in turn configures the VSN of the tenant.

In this paper, Section II motivates the research and Section III covers SDSN. Section IV presents our FM4SN approach in detail, and Section V evaluates it. Section VI summarizes the related work, and Section VII concludes the paper.
\section{Motivation and General Requirements}
RoSAS (Road-Side Assistance Service) offers roadside assistance to its tenants such as travel agencies and vehicle sellers by using services such as repairers and tow trucks (Fig 1). RoSAS adopts the SIMT model for its service network. Each tenant has a virtual service network to coordinate assistance for their users such as travelers and motorists.
\begin{figure}
\centering
\includegraphics[width=\columnwidth]{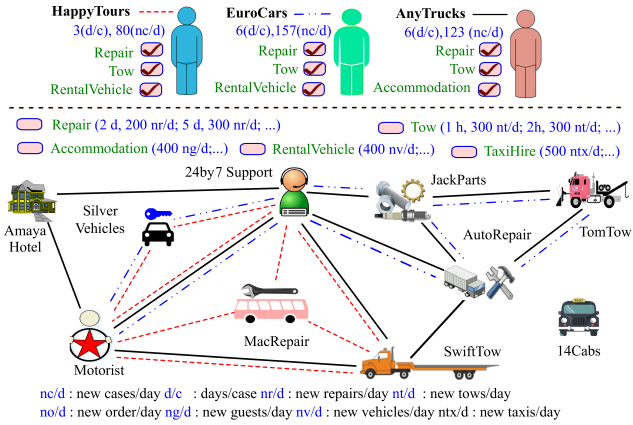}
\caption{RoSAS service network, its partner organizations, and its tenants}
\end{figure}

To execute the business processes in the service network, the software engineers at RoSAS design, enact, and manage it using a software framework (e.g., SDSN \cite{R3}). For a new tenant, a software engineer creates a VSN in the service network to meet the functional and performance requirements of the tenant. As these requirements change, the software engineer reconfigures the VSN. To reduce this operational overhead, RoSAS wants the tenants to create and reconfigure their own VSNs. However, the customization options and their dependencies can be numerous and complex, and the customization activities need platform-specific knowledge, which are too complex for the tenants to perform directly. To alleviate these complexities, RoSAS needs a customization support. Consider three key requirements for such support. 
\subsubsection{A High-Level Customization Support} The tenants should be guided to select or deselect the desired and allowed customization options of the service network according to their functional and performance needs. Such options should represent high-level end-user visible capabilities of the service network to reduce the customization complexity. For example, the tenant HappyTours should be able to select repair, vehicle renting, and towing capabilities, and choose the performance options for repair as 2 days of repair duration and 80 new repairs per days. HappyTours should be also able to check if the selected capabilities and their performance options are compatible, and can be provided by the service network, and change the selection as necessary. 
\subsubsection{Automated Provisioning} A middleware at RoSAS hosts the composite application that realizes the roadside assistance service network. Upon receiving the requirements of a tenant, the middleware should automatically configure a VSN for the tenant in the service network to realize the requirements. For example, a VSN for HappyTours should realize the repairing, vehicle renting, and towing capabilities and the selected performance options for these capabilities. 
\subsubsection{Configuration and Reconfiguration} The tenants should be able to create and reconfigure their own VSNs of the RoSAS service network to meet their initial and changing business needs. For example, after one year, HappyTours wants to use accommodation instead of vehicle rental, and to reduce the number of the repairs per day from 80 to 70. The VSN of HappyTours needs to be reconfigured accordingly. 

\section{Overview of Multi-tenant Service Networks}
We realize multi-tenant service networks with the SDSN \cite{R3}, and present an overview of it as the basis of understanding our FM4SM approach (see \cite{R3} for more details).

A multi-tenant service network has four layers (see Fig. 2): partner services, service network, virtual service networks (VSNs), and control/management plane. The topology of the service network acts as an overlay network over the services and makes the message flows between them explicit and structured. A \textit{node} in the topology is a proxy to a service, and can intercept and route messages from/to the service. A \textit{contract} between two nodes models the contractual relationships between the corresponding services, and acts as the messaging channel between the nodes. 

RoSAS service network in Fig 2 consists of a number of nodes (e.g., MO, SC, and GR1) connected by contracts (e.g., MO-GR1 and MO-SC), and supports the coordination of the services (e.g., motorist, 24by7, and MacRepair) to meet the roadside assistance requirements of the tenants. 
\begin{figure} [!t]
\centering
\includegraphics[width=\columnwidth]{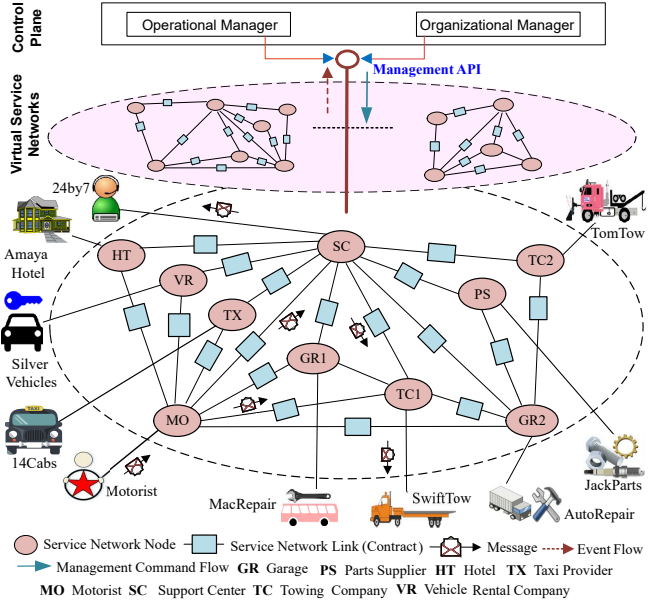}
\caption{Multi-tenant service networks with SDSN}
\end{figure}
\begin{figure}[!t]
\centering
\includegraphics[width=0.85\columnwidth]{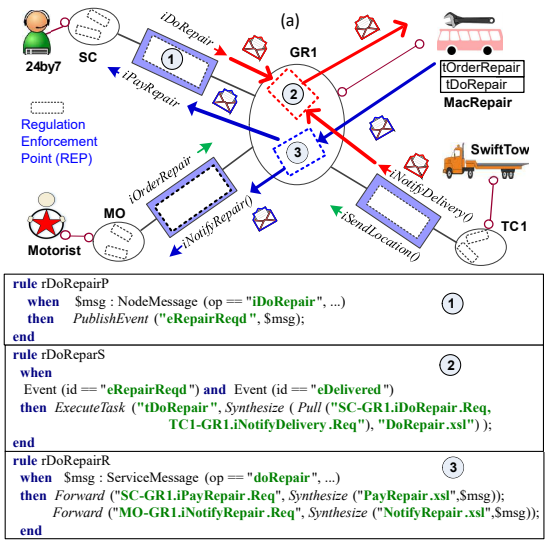}
\caption{Repair collaboration with MacRepair, REPs and some rules}
\end{figure}

The services in the service network in collaboration realize the requirements of the tenants. \textit{A collaboration} involves a subset of interacting services in the service network, and realizes a particular user/tenant requirement. It has a topology (\textit{configuration design}) and the interaction messages between the services of the collaboration pass over this topology. The topology has a set of \textit{regulation enforcement points (REPs)} (at each node and contract) to intercept and regulate message exchanges between services over it (\textit{regulation design}). Each REP has a \textit{knowledgebase} and a \textit{regulation table}. The former contains rules that implement regulation decisions using \textit{regulation mechanisms} such as admission control and message transformation. The latter maps a message flow to a set of rules in the knowledgebase, which decide what to do with the message flow.

Fig 3 shows the repair collaboration with MacRepair. Its topology consists of the nodes GR1, MO, SC, and TC1 and their services. The rules at the REPs over the topology regulate the message passing over it. A rule (1) at the contract SC-GR1 processes the interaction \textit{iDoRepair}, generating event \textit{eRepairReqd}. Through the relevant events, a rule (2) at the node GR1 synchronizes over the interactions \textit{iDoRepair} and \textit{iNotifyDelivery} from the nodes SC and TC1 to invoke the task \textit{tDoRepair}, which sends a request to MacRepair. A rule (3) at the node GR1 routes the response from MacRepair to the nodes SC and MO as the interactions \textit{iPayRepair} and \textit{iNotifyRepair} via their contracts. 

A VSN represents a specific service composition in the service network that meets the functional and performance requirements of a tenant. The topology of a VSN is formed and reconfigured by adding and changing the entries to the regulation tables at REPs. A table entry maps the messages belonging to a VSN to a subset of the rules at the REP. These rules intercept, route, and regulate of the messages over the topology of the VSN. The execution of the VSN for a request from a user of the tenant represents a VSN instance. The similarities and differences in functional and performance needs drive the sharing and variations in topologies of VSNs and regulation rules used (see the VSNs in Fig 2). 

A VSN composes a subset of collaborations. A VSN includes (excludes) a collaboration in its topology by adding (removing) the regulation table entries required to enforce the configuration and regulation designs of the collaboration. For example, the VSN \textit{HappyTours} can include the repair collaboration with MacRepair by adding the table entries relevant for the collaboration (e.g., \textit{(HappyTours, \{rDoRepairP, ...\})} to the REP of the contract SC-GR1 and \textit{(HappyTours,\{rDoReparS, ...\})} to the REP of node GR1).

Analogous to the data and control plane separation in software-defined networking (SDN) \cite{R5}, in a multi-tenant service network, the services and the service network (with VSNs) that connects them constitute the data plane, while the management logic of a service network resides mainly in the control plane (the top layer of Fig 2). A service network offers the management interface that can be used to monitor it, and to alter its topology (add, remove, and update nodes and contracts) and its regulation structure (add, remove, update regulation rules and table entries). The organizational manager and the operational manager in the control plane use these interfaces to enforce management policies.

\section{FM4SN : A Feature-Oriented Customization Approach for Multi-Tenant Service Networks}
Our FM4SN (Feature Models for Service Networks) approach enables the tenant-driven customization of single-instance multi-tenant service networks. This section describes the FM4SN approach in detail.

\subsection{Architecture of FM4SN}
\begin{figure}[!t]
\centering
\includegraphics[width=\columnwidth]{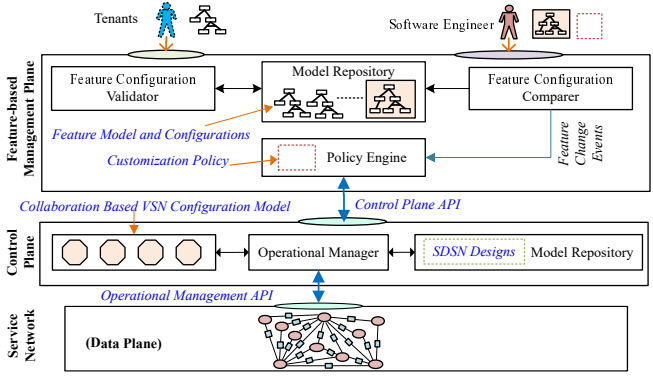}
\caption{Architecture of FM4SN}
\end{figure}
Fig 4 shows the high-level architecture of FM4SN, which extends the SDSN architecture by enhancing the control plane layer and introducing a new high-level management layer. The control plane provides a collaboration based abstraction to specify VSNs (virtual service networks). Via the control plane API (analogous to the northbound interfaces of SDN), a VSN can be configured by including or excluding collaborations available in the service network. The design models of the service network stored in the model repository include the information about the collaborations.

To reduce the complexity of creating and customizing their own VSNs by the tenants, the high-level management plane provides a feature-based configuration abstraction. We use the feature model from software product lines \cite{R13, R14}. A feature model can represent the functional and performance requirement options, their inter-dependencies, and guide the tenants to select a valid and desired set of options (a valid feature configuration) for their VSNs. The management plane uses an ECA (event-condition-action) rule based customization policy for specifying the realization of the features in the service network, in terms of feature to collaboration mappings. The events are feature-level customization events, and the actions are the control plane API operations. For a valid feature configuration from a tenant (a customization request), the feature configuration comparer generates customization events by comparing it with the current feature configuration (if there is any) and the feature model. These events trigger the rules in the customization policy, which in turn execute the control plane API operations, and consequently configure the routing logic in the service network to enforce the new or changed VSN. 
\subsection{Feature-based Configuration Model}
We use feature models \cite{R13,R14} to capture the commonality and variability in tenants’ customization requirements at a high-level of abstraction. A valid configuration of the feature model (\textit{feature configuration}) represents a VSN as a set of functional and non-functional features. 

\begin{figure*}[!t]
\centering
\includegraphics[width=\textwidth,keepaspectratio]{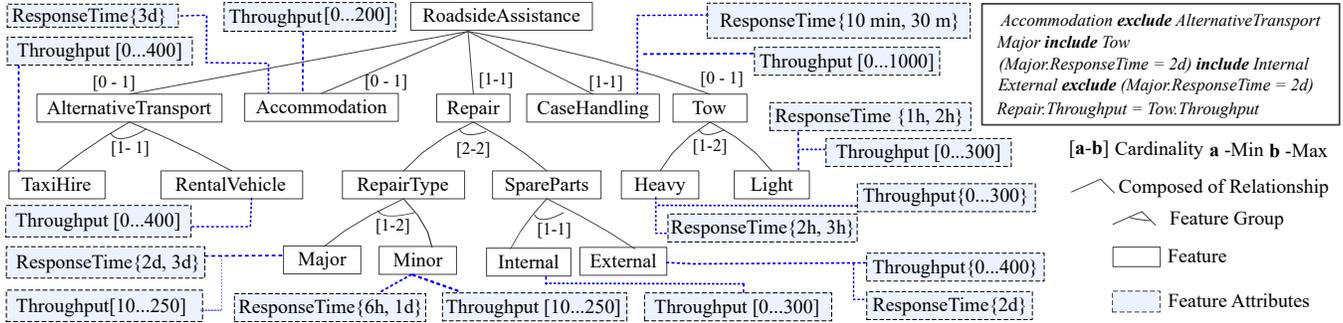}
\caption{The feature model for the RoSAS service network}
\end{figure*}

A feature can represent a distinctive functional requirement, behavior, or characteristic of one or more VSNs. A requirement option may only be used by a subset of tenants (\textit{optional}), or must be used by each tenant (\textit{mandatory}), and the cardinality of the corresponding feature node can specify these constraints. A tenant may select a cohesive subset of requirement options, and the group cardinality of a parent feature node can specify this constraint.

Consider the feature model for the roadside assistance service network (Fig 5). The cardinality \textit{[1-1]} of the feature \textit{Repair} implies that it is mandatory, and that \textit{[0-1]} of the feature \textit{Tow} indicates that it is optional. The group cardinality \textit{[1-2]} of the feature \textit{RepairType} implies that at least one of its two children features \textit{Major} and \textit{Minor} must be selected (\textit{Or}), and that \textit{[1-1]} of the feature \textit{AlternativeTransport} indicates that its children features \textit{TaxiHire} and \textit{RentalVehicle} are alternatives. The group cardinality \textit{[2-2]} of the feature \textit{Repair} specifies that both of its children features \textit{RepairType} and \textit{SpareParts} must be selected (\textit{And}).

A performance option for a functional requirement consists of the values for one or more performance parameters of the functional requirement. The attributes of a feature can model the performance parameters for a functional requirement \cite{R14}. In Fig 5, the attributes of the feature \textit{Major} specifies the performance of the capability major repair. Its response time can be one of \{2 days, 3 days\}, and its throughput can range from 10 to 250 new repairs per day.

We codify the constraints on requirement options using the feature model constraints, which can define the dependencies (e.g., include/exclude) between features, between features and attributes, and between attributes. In general, the types of constraints and analysis for a feature model are only limited by the underlying reasoning language and model such as constraint logic and propositional logic \cite{R14}. A valid feature configuration respects these constraints and those of the meta-model of the feature model (i.e., mandatory, optional, feature cardinality, and group cardinality)

Consider some of the constraints on the roadside assistance requirements (Fig 5). Although a minor repair can be done on-site, a major repair needs the towing of the broken-down vehicle to a garage. Thus, the tenants who need the support for a major repair must also use towing. This constraint can be specified as \textit{Major include Tow}. If some tenants prefer to use external spare parts, then they cannot select the response time of 2 days, as only AutoRepair uses external spare parts and its average response time is 3 days. The corresponding constraint in the feature model is \textit{External exclude (Major.ResponseTime = 2d)}.

\subsection{Mappings from Feature Model to Service Network}
Table 1 shows the relationships between features and their realization in the service network. In general, \textit{a collaboration} among a set of services realizes a high-level functional requirement or feature \cite{R3, R15}. The collaborations with the variations in services and their capabilities and performance, configuration design (topology and control flow), and regulation design realizes performance variations for a functional feature, resulting a set of alternative collaborations.
\begin{table}[!t]
\centering
\caption{Feature-based model to service network model mappings }
\begin{adjustbox}{max width=0.86\columnwidth}
\begin{tabular}{|l|l|}
\hline
Feature                                                                         & Service Collaboration                                                                                                              \\ \hline
Feature Composition                                                             & Collaboration Composition                                                                                                          \\ \hline
Feature Dependency                                                              & Collaboration Dependency                                                                                                           \\ \hline
\begin{tabular}[c]{@{}l@{}}Performance Variations \\ for a Feature\end{tabular} & \begin{tabular}[c]{@{}l@{}}Variations in Services, Configuration Design\\  and Regulation Design of the Collaboration\end{tabular} \\ \hline
Feature Model                                                                   & Service Network                                                                                                                    \\ \hline
Feature Configuration                                                           & Virtual Service Network                                                                                                            \\ \hline
\end{tabular}
\end{adjustbox}
\end{table}

A collaboration between MacRepair, Motorist, and SupportCenter (Fig 3) can provide the \textit{Repair} feature. A collaboration between AutoRepair, Motorist, SupportCenter, and Spare Parts Supplier can also realize the same feature. The capabilities and performance properties of MacRepair and AutoRepair are different (Fig 1). The regulations applied to the message exchanges between the services can also have variations, reflecting the differences in the contractual agreements between services. The variations in services, configuration topologies and control flows, and regulations applied in these collaborations make them suited to support different performance options (2 days vs 5 days) for \textit{Repair}.

A feature hierarchy can be used to refine a feature in groups of increasing levels of detail. This refinement can manifest as the constraints on the selection of specific collaboration instances. For example, the feature \textit{Repair} can be realized using a repair collaboration with AutoRepair or that with MacRepair. This feature can be refined by specifying the selection of spare parts (\textit{Internal} or \textit{External} children features). As AutoRepair does not have internal spare parts (service capability differences), the corresponding collaboration cannot support \textit{Repair} with \textit{Internal} option.

To compose two features, the corresponding collaborations need to be merged together. The two collaborations can be overlapping or disjoint.  In the first case, the union of the regulation table entries for the collaborations realize the merged collaboration. In the second case, a subset of interactions produced by one collaboration will be consumed by other collaborations, and vice versa. To capture the routing and regulation of these inter-collaboration interactions, we use the concept (abstraction) of inter-collaboration regulation units (as in SDSN \cite{R3}). Similar to the realization of a collaboration, these units are also manifested in the service network as a set of regulation table entries. Consider the composition of the repair collaboration with MacRepair and the towing collaboration with SwiftTow. SwiftTow delivers a broken-down vehicle to the location given by MacRepair, which requires routing of the interaction messages \textit{iSendLocation} and \textit{iNotifyDelivery} between two services. An inter-collaboration regulation unit (\textit{SwiftTowAndMacRepair}) can represent the relevant routing logic.

The dependencies and constraints between features can drive inclusion and exclusion of the collaborations that realize the features in a VSN. For example, the constraint \textit{Major include Tow} can be enforced by including a towing collaboration in the VSN along with a repair collaboration.

A service network is designed to realize the functional and performance requirements of all possible tenants, which are captured by the feature model. A tenant selects a valid configuration of the feature model as a set of functional and performance features. A VSN in the service network realizes the selected feature set by composing the regulated collaborations that implement those features.
\subsection{Designing Customization Policies}
To allow a software engineer to design the customization logic for a multi-tenant service network, we provide an ECA rule based language that uses the mappings between features and collaborations. Fig 6(a) shows its key concepts.
\begin{figure}[!t]
\centering
\includegraphics[width=\columnwidth]{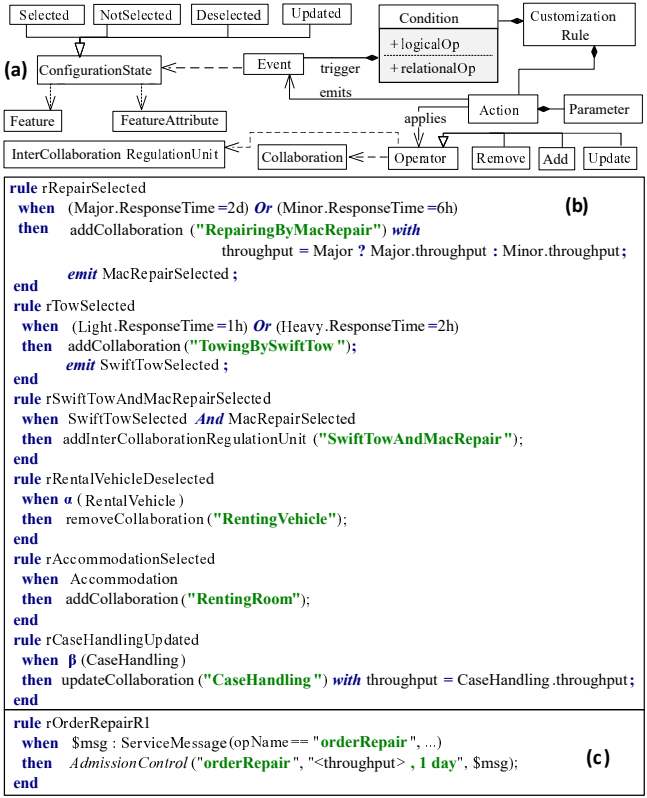}
\caption{Customization policy language (a) meta-model, (b) an example, (c) a parameterized regulation rule }
\end{figure}

A \textit{condition} of a rule is a logical expression of events. A customization event captures the state of a feature/attribute, which are fourfold: \textit{Selected}, \textit{NotSelected}, \textit{Deselected}, and \textit{Updated}. The actions include \textit{Add}, \textit{Remove}, and \textit{Update} operations on collaborations and inter-collaboration regulation units. The control plane of the service network implements these operations using the operational management interface of the service network, which support adding, removing, updating regulation table entries and rules at REPs. To add a collaboration to a VSN, the entries (the VSN identifier to regulation rule mappings) are added to the regulation tables at the relevant REPs over the topology of the collaboration. The removal of a collaboration from a VSN removes the added entries. The update operation can reinitialize the parameterized rules used by the collaboration.

The regulation rules use regulation mechanisms to implement regulation decisions. The parameters for these mechanisms need to be able to be provided, per tenant, as necessary. For this purpose, we use the \textit{parameterized} rules, whose parameters are mapped to feature attributes. 

There exist dependencies between customization decisions. An enactment of a given customization decision may require the enactment or prevention or revocation of some other customization decisions. To capture these dependencies, we introduce an action to generate custom events. A rule can emit an event indicating the state (e.g., completion) of the enactment of a customization decision. The dependent rules can use that event in their conditions.

Fig 6(b) shows a fragment of the customization policy for RoSAS service network. The symbols \(\sim (not)\), \(\alpha (alpha)\), \(\beta (beta)\) represent the events \textit{NotSelected}, \textit{DeSelected}, and \textit{Updated}, respectively. A VSN can include the collaboration \textit{RepairingByMacRepair} to support major repairs (2 days repair duration) or minor repairs (6 hours repair duration). The number of repairs should be limited to the value specified (throughput) by the tenant. The rule \textit{rRepairSelected} defines this customization decision. The required parameterized rule is in Fig 6(c). If the collaborations \textit{RepairingByMacRepair} and \textit{TowingBySwiftTow} are included in a VSN, an inter-collaboration regulation unit should also be included to wire them. The rule \textit{rSwiftTowAndMacRepairSelected} captures this decision, and is triggered once the rules for adding those two collaborations are enacted (event-based dependencies). 
\subsection{Configuration and Reconfiguration of VSNs}
The tenants can retrieve the feature model of the service network via the Web service interface of the high-level management plane, and can create a feature configuration using the feature modelling tools \cite{R14}. They can check a feature configuration for its validity and its completeness using the rich support for feature model analysis \cite{R14}.

Upon receiving a feature configuration, the management plane first validates it against the feature model. An invalid configuration is rejected, and a valid configuration is used to generate customization events, reflecting the states of each feature and attribute. In the configuration phase, the states can be \textit{Selected} or \textit{NotSelected}. The policy engine executes the customization policy with these events. The policy rules are triggered as their conditions are satisfied, and the desired changes are propagated to the service network.

Consider the provisioning of a VSN for HappyTours. The feature configuration includes the features such as \textit{Major (ResponseTime=2d)}, \textit{Heavy (ResponseTime=2h)}, \textit{RentalVehicle}, \textit{CaseHandling (Throughput=80)}. This configuration generates \textit{Selected} events for features and attributes in the configuration (e.g., \textit{Major} and \textit{Major.ResponseTime=2d}) and \textit{NotSelected} events for those not in the configuration (e.g., \textit{Accommodation} and \textit{Minor.ResponseTime}). These events trigger the rules in the customization policy, for example, rules \textit{rRepairSelected} and \textit{rTowSelected} in Fig 6(b). 

To reconfigure a VSN, a tenant can retrieve the feature configuration of the VSN and update it to match the changed requirements. Compared with the configuration phase, the management plane performs similar activities. However, the state of a feature or an attribute can be \textit{Selected}, \textit{NotSelected}, \textit{Updated}, and \textit{DeSelected}. The last two state changes are determined by comparing the new configuration and the current configuration. A change to the selection of an attribute or its value is considered as an update to the feature.

Suppose HappyTours prefers to use accommodation, replacing rental vehicle, and to reduce the number of cases per day by 10. To express these requirement changes, HappyTours can reconfigure its feature configuration by selecting \textit{Accommodation}, deselecting \textit{RentalVehicle}, and updating the throughput attribute of \textit{CaseHandling}. This configuration will result in the events \textit{DeSelected} for \textit{RentalVehicle}, \textit{Selected} for \textit{Accommodation}, and \textit{Updated} for \textit{CaseHandling} and \textit{CaseHandling.throughput}. These events trigger the rules such as \textit{rRentalVehicleDeselected}, \textit{rAccommodationSelected}, and \textit{rCaseHandlingUpdated} to reconfigure the VSN (Fig 6).
\section{Prototype and Evaluation}
\textbf{Prototype.} We extended SDSN framework\cite{R3}. We have implemented and added the control and management planes as modules to Apache Axis2 Web service engine (axis.apache.org/axis2). The management interfaces of each plane is exposed as Axis2 Web services. We used the FAMA framework (www.isa.us.es/fama/)\cite{R14} for analyzing feature models. We implemented the domain specific customization policy language (design) using the Xtext framework (www.eclipse.org/Xtext/). We used Drools rule engine and its language (www.drools.org) to define and enact the executable policies. The source code of the prototype has been integrated to the GitHub repository of the SDSN (https://github.com/road-framework/SDSN).

\textbf{Case Studies.} We have fully developed the roadside assistance case study presented in this paper, and a camping assistance case study. The features of the roadside assistance are \textit{CaseHandling}, \textit{Reimbursement}, \textit{Tow}, \textit{Repair}, \textit{SpareParts}, \textit{VehicleRepairAssessment}, \textit{LegalAssistance}, \textit{Accommodation}, \textit{RentalVehicle}, and \textit{TaxiHire}. Those for the camping assistance are \textit{TourArrangement}, \textit{EquipmentRental}, \textit{TaxiHire}, \textit{CaseHandling}, \textit{BikeRental}, and \textit{GroceryDelivery}. Each feature had one or more performance options. A collaboration among a set of services realizes a feature. Each case study had three tenants with different functional and performance requirements (three different feature configurations). We validated the provisioning of a VSN for a feature configuration by comparing the logs and the response messages of the VSN execution with those of the manually created VSNs. The case study resources are at https://github.com/scc2018fm4sn/FM4SNCaseStudy.

\textbf{Performance Evaluation.} Our experiment setup comprises a workload generator (Apache AB), the middleware (SDSN with our customization module), and services (Axis2). The experiment used a single machine with an Intel i7-7700HQ CPU @ 2.80GHz processor with 16GB RAM and Windows10 64bit. We quantify the overhead of the customization, which is the time difference between the service network manager receiving a customization request and the VSN being ready for use after applying the customization policy. To measure the overhead for the worst case scenario, we use a VSN that covers the entire service network and uses all functional features. We send the customization requests to create the VSNs using the Web service API of the service network manager. Fig 7 shows the variation of this overhead with the number of the customization requests for both case studies. The overhead increases approximately linearly.
\begin{figure}
\centering
\includegraphics[width=0.8\columnwidth]{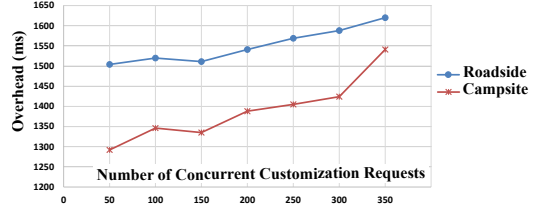}
\caption{Customization overhead (milliseconds)}
\end{figure}
\section{Related Work}
The approaches to customizing composite applications \cite{R6, R7, R8} can generate and reconfigure application variants for users considering both functional and non-functional requirements. In general, these works use a model-driven approach. The commonality and the variability at the user requirements and the composite service application are modeled and mapped. A variant of the composite application is generated for a variant of the requirement variability model. 

The approaches to customizing multi-tenant applications \cite{R9,R10,R11,R12,R17} also use a model-driven approach. Their capabilities include modeling the variability in requirements and composite applications, generating application variants automatically or semi-automatically for a set of requirement options, and reconfiguring the variants \cite{R12,R17,R18}.

The service network model can provide a natural abstraction to design, enact, regulate, and manage webs of real-world business service networks \cite{R1, R2, R3}. Most existing studies consider the modeling and analysis of service networks from specific aspects such service relationships \cite{R1} and change impacts \cite{R19}. SDSN \cite{R3} supports the design, enactment, and evolution of service networks. Recently, the service mesh \cite{R20} was used to describe the inter-service communication infrastructure for microservices. The service mesh (the data plane) has a control plane to monitor and manage it. Istio \cite{R20} supports the MIMT model (one control plane and one mesh per tenant) for the service mesh.

These related studies have several limitations. First, there is no high-level customization support for (multi-tenant) service networks. The customization support available for process-centric models (e.g., BPMN) or component-centric models (e.g., SCA) cannot be directly used for the service network model as those models (including service mesh) cannot represent the domain concepts (e.g., services, service capabilities, service relationships, service interactions, interaction routing and regulation, and virtualization) of a service network adequately. The network view can improve the flexibility to form virtual application variants on demand (as dynamic routing based virtualization of computer networks with SDN). These application variant can have different topologies, conversational behaviors, and regulation applied to such behaviors to support performance variations. Second, most studies only support the MIMT model as they create and maintain physically separated variants. A few works that support the SIMT model only consider functional requirements. Some of them focus on component-based applications. Finally, most assume a feature is mapped to a service, a component, or an arbitrary application fragment. There is the lack of a natural correspondence between domain requirement options and realization options, resulting in complexity in the traceability between them, in the (re)configuration activities, and in the associated models.

To alleviate the aforementioned limitations of the related work, we have presented a tenant-driven customization approach for SIMT service networks. By adopting a network view (controlled message passing between entities), we model the (re)configuration of a composite application as changing the routing and regulation of the message passing over the runtime architectural model of the application realized as a service network. The tenants are provided a feature-based high-level model to hide the complexity of creating and configuring virtual service networks (VSNs) or application variants. To modularize the changes to the routing and regulation logic, we use \textit{a regulated collaboration} among a set of services to realize a feature. The customization policy at the service network manager enforces the feature-level customization requests from the tenants. The customization policy language utilizes the concepts of features and collaborations, and the mappings between them.
\section{Conclusion}
We have introduced a tenant-driven customization approach for multi-tenant service networks. A tenant can specify its functional and performance needs by selecting and deselecting high-level configuration options (features). The high-level requirements of a particular tenant is realized by a virtual service network (VSN) on the same service network shared with other tenants. The manager of the service network automatically enacts high-level (re)configurations by (re)configuring VSNs at runtime. We have discussed the architecture of the customization support, feature-based configuration model, customization policy, and (re)configuration processes. We have evaluated our approach with two case studies and a related analysis on customization overhead. 

We are currently exploring the use of the service network model (as an alternative to the service mesh model) in composing, virtualizing, and managing microservices.

\section*{Acknowledgments}
This research is partially supported by the European Commission grant no. 825480 (H2020), SODALITE, and by Swinburne University of Technology.

\end{document}